\documentclass[conference]{IEEEtran}
\usepackage{url}
\usepackage{subcaption}

\usepackage[breaklinks]{hyperref}

   \usepackage{graphicx}
   \graphicspath{figures/}
   \DeclareGraphicsExtensions{.pdf,.jpeg,.png}

\usepackage{booktabs}
\usepackage{comment}
\usepackage[normalem]{ulem}
\usepackage{multirow}
\usepackage{float}

\usepackage{listings}
\usepackage{xcolor}
\usepackage{hyperref}

\newcommand\thename{\textsc{Sn4ke}}
\usepackage{multirow}
\newcommand{\translate}[0]{\begin{small} {\tt translate}\end{small}}
\newcommand{\smallf}[1]{\begin{small}#1\end{small}}

\definecolor{codegreen}{rgb}{0,0.6,0}
\definecolor{codegray}{rgb}{0.5,0.5,0.5}
\definecolor{codepurple}{rgb}{0.58,0,0.82}
\definecolor{backcolour}{rgb}{0.95,0.95,0.92}
 
\lstdefinestyle{mystyle}{
    backgroundcolor=\color{backcolour},   
    commentstyle=\color{codegreen},
    keywordstyle=\color{magenta},
    numberstyle=\tiny\color{codegray},
    stringstyle=\color{codepurple},
    basicstyle=\ttfamily\footnotesize,
    breakatwhitespace=false,         
    breaklines=true,                 
    captionpos=b,                    
    keepspaces=true,                 
    numbers=left,                    
    numbersep=5pt,                  
    showspaces=false,                
    showstringspaces=false,
    showtabs=false,                  
    tabsize=2
}
 
\begin{document}
\lstset{language=C}

%
\title{\thename{}: Practical Mutation Testing at Binary Level}

\author{
\IEEEauthorblockN{Mohsen Ahmadi}
\IEEEauthorblockA{Arizona State University\\
pwnslinger@asu.edu}
\and
\IEEEauthorblockN{Pantea Kiaei}
\IEEEauthorblockA{Worcester Polytechnic Institute\\
pkiaei@wpi.edu}
\and
\IEEEauthorblockN{Navid Emamdoost}
\IEEEauthorblockA{University of Minnesota\\
navid@cs.umn.edu}

}

\IEEEoverridecommandlockouts
\makeatletter\def\@IEEEpubidpullup{6.5\baselineskip}\makeatother
\IEEEpubid{\parbox{\columnwidth}{
    Workshop on Binary Analysis Research (BAR) 2021 \\
    21 February 2021, Virtual\\
    ISBN 1-891562-62-2 \\
    https://dx.doi.org/10.14722/bar.2021.xxxxx\\
    www.ndss-symposium.org
}
\hspace{\columnsep}\makebox[\columnwidth]{}}


\maketitle

\begin{abstract}
Mutation analysis is an effective technique to evaluate a test suite adequacy in terms of revealing unforeseen bugs in software. Traditional source- or IR-level mutation analysis is not applicable to the software only available in binary format. This paper proposes a practical binary mutation analysis via binary rewriting, along with a rich set of mutation operators to represent more realistic bugs. We implemented our approach using two state-of-the-art binary rewriting tools and evaluated its effectiveness and scalability by applying them to SPEC CPU benchmarks. Our analysis revealed that the richer mutation operators contribute to generating more diverse mutants, which, compared to previous works leads to a higher mutation score for the test harness. We also conclude that the reassembleable disassembly rewriting yields better scalability in comparison to lifting to an intermediate representation and performing a full translation. 

\end{abstract}


%

\section{Introduction}
\label{intro}



In many contexts software is accompanied with test suites when delivered to the customer. In such cases, the test suite is the primary and mostly the only leverage to assure correctness of the software. This demonstrates the importance of test suite adequacy. Evaluating such adequacy requires a separate analysis which is known as mutation analysis. Such analysis generates \textit{mutants} of the original program via making small changes either at source or IR level. Such mutants then are ran through the test suite. If the test suite successfully differentiates a mutant from the original program based on its observed output, it is said that it has \textit{killed} that mutant. The ratio of killed mutants to all tried mutants is called \textit{mutation score} which is the metric to describe the adequacy of the test suite.

In order to create mutants, many researches proposed various mutation operators which are the rules for changing statements or instructions in the original program to get a new mutant. A good mutation operator is representative of a real world bug that may be introduced as a result of errors either in development or the building process. For example, consider a bug-fixing patch which is accompanied with a test. A mutation operator to evaluate the test’s adequacy would be undoing the effect of the patch, e.g if the patch introduces a bounds check, the mutation operator causes skipping the check.

In many scenarios the program source code is not available like in proprietary software or shared libraries. In these cases the mutation tool should still be able to generate mutants from the original binary. Binary level mutation has been explored previously \cite{emamdoost2019binary,BeckerBinMutation}. The major challenge on binary mutation tools is practical and scalable binary rewriting which in turn imposes limitation on the binary mutation operators. In order to generate mutants from the original binary, the mutation tool needs to restore a higher level code representation like assembly and then apply the mutation operator which may change the instructions address layout. Binary rewriting has found many applications in program analysis, security and etc. Mutation analysis is another application of binary rewriting which requires generation working mutants by making changes at instruction level.

Binary rewriting techniques have been extensively incorporated in security enforcement tools like ~\cite{mccamant2006evaluating, yee2009native,zhang2013control,emamdoost2018sfi} where the instructions are aligned, inserted or replaced to enforce measurements like control flow integrity or fault isolation and etc. In recent years reliable binary rewriting techniques (like Ddisasm~\cite{flores2020datalog} or Ramblr~\cite{wang2017ramblr})  have been proposed which expand the domain of such measurements applications to the programs available only in binary format. 
Previously researchers have demonstrated the applicability of reassembleable disassembly for binary mutation \cite{emamdoost2019binary}. They used Uroboros~\cite{wang2016uroboros} and were able to mutate binaries from SPEC CPU. Limitations imposed by Uroboros obstructs the practicality of such application, more specifically authors reported fragility of the tool in a way that it worked for specific compilation options. Their implemented mutation operators are limited to only conditional jump and move instructions. 

In this paper, we revisit the notion of binary mutation in light of recent advancements in binary rewriting to implement a practical binary mutation tool that can support real-world binaries. Additionally, we employ a richer set of mutation operators that span over conditional, logical, and arithmetic instructions in order to have a more rigorous evaluation of the test harness. Our binary mutation tool named \thename{} is accessible at \url{https://github.com/pwnslinger/sn4ke/}.

The rest of the paper is as follows: in \autoref{sec:related} we discuss the related work on the subject of binary mutation analysis of tests. \autoref{sec:approach} presents our approach and design of \thename{}, specifically the set of mutation operators and the binary rewriting engine. \autoref{sec:eval} describes our evaluation of \thename{} on SPEC CPU benchmarks and the comparison of performance of the two binary rewriting tools we used. \autoref{sec:discuss} discusses some of interesting challenges we had in adapting Rev.ng for our purpose. Finally, \autoref{sec:conclude} concludes the paper.

\section{Related Work}
\label{sec:related}

\begin{table*}
  \caption{Mutation Operator Classes}
  \centering
\begin{tabular}{ll}
\hline \toprule
Mutation Class              & Description        \\
\midrule
\midrule
\multirow{2}{*}{Arithmetic} & Replace arithmetic assignment operators from the set of \{+=, -+, *=, /=\}           \\
                            & Replace with an operator from the set of \{+, -, *, /, \%\}                          \\
\midrule
\multirow{3}{*}{Logical}           & Substitute with another bit-wise logical operator from \{\textasciicircum{}, $\mid$, \&\} \\
                            & replace with a logical assignment from \{\textasciicircum{}=, $\mid$=, \&=\}              \\
                            & Substitute the connector with another logical operator from \{\&\&, $\parallel$\}   \\
\midrule
\multirow{2}{*}{Conditional}           & Substitute any conditional jump with an unconditional branch to force taking the branch \\
                            & taking the fall-through (instruction following the branch) edge of branch by NOPing the condition   \\
\midrule
Constants           &   Replace any immediate value \textit{c} with one another constant from set \{-1, 0, 1, -c, c+1, c-1\} \\
\midrule
Skip                &   Replace instructions from set Arithmetic, Logical, and Conditional with a NOP operator to skip the execution of that operator \\
\bottomrule
\end{tabular}

  \label{tab:mutation-operators}
\end{table*}

Generating variants of a program has been employed in the context of security: fuzzing tools like T-Fuzz~\cite{tfuzz} negate the conditions to pass the blocking conditions and explore new paths; additionally, tools like EvilCoder~\cite{evilcoder} LAVA~\cite{lava} insert exploitable bugs in the program to provide testing corpus for vulnerability detection tools. In a broader context, mutation analysis tends to generate variations of the software and use it as a data set to evaluate the adequacy of tests to catch such variants. Authors in \cite{emamdoost2019binary} employed Uroboros \cite{wang2016uroboros}, a reassembleable disassembly binary rewriting tool to recover the assembly representation of a binary and apply the mutation at the assembly level. They applied a limited set of mutation operators mainly focusing on flag-use instructions and used the generated mutants to evaluate the adequacy of tests for SPEC CPU benchmarks. Their findings demonstrates that the tests designed for performance benchmarking catch fewer mutants compared to tests designed for code coverage. 

As we are focused on mutation analysis at binary level, therefore we describe works related to mutation analysis and binary rewriting in the following subsections. 

\begin{figure*}
\centering
\includegraphics[width=.95\textwidth]{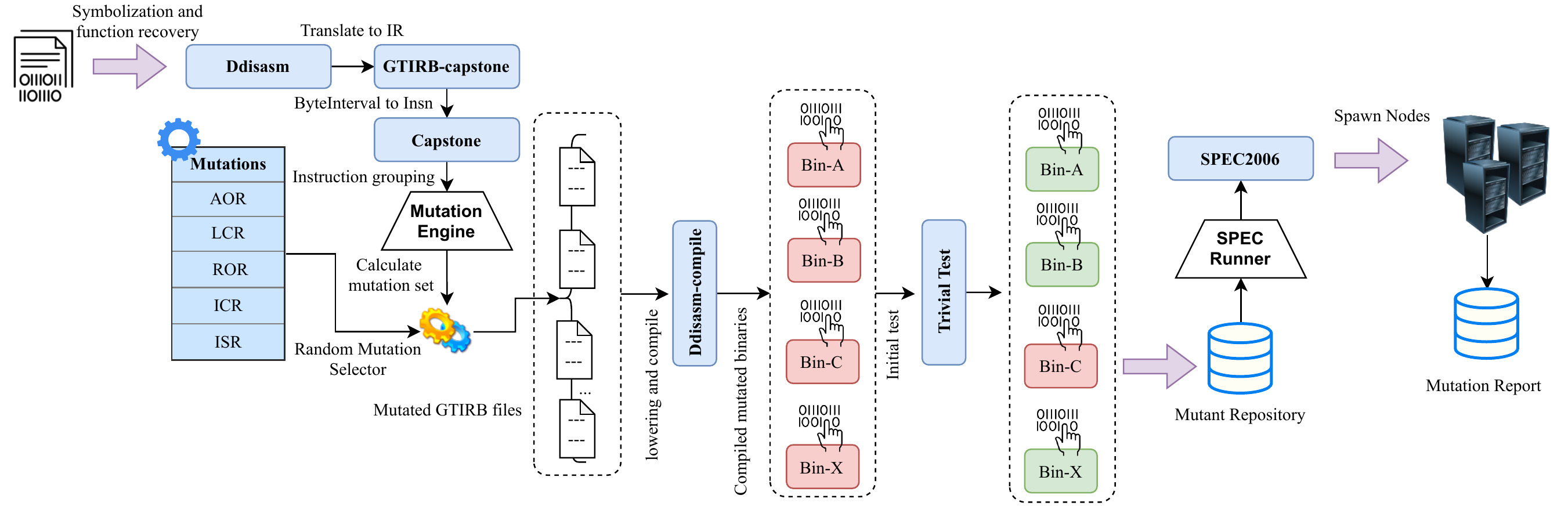}

\caption{
\thename{} Workflow of our binary rewriter: First we pass the binary under test to Ddisasm for reconstructing the relocation table by identifying the instruction boundaries using backward and forward traversal techniques described in their paper \cite{budd1978design}. As a result we can retrieve the lifted binary in GTIRB representation which contains symbolized addresses for references and variables. Gtrib-capstone is the built-in rewriter and we modified the code to fix some bugs related to shifting the symbolic expression of blocks following the insertion point. Gtirb represents binary as a module contains functions and each function consists of various blocks. Blocks can either be DataBlock or \textit{CodeBlock} objects. Each block encapsulates \textit{ByteIntervals} which is a direct mapping of gtirb IR to assembly instructions. Next, mutation engine finds the matching operators and calculates all possible mutations fall under any of mutation class categories. Then by applying each mutation resulting Gtirb file passed to Ddisasm-compiler to re-compile the disassembly. Resulting binaries should pass initial tests before sending out to the SPEC Runner. 
\label{fig:workflow}}

\end{figure*}

\subsection{Mutation Analysis}
Since its introduction \cite{budd1978design,DeMilloTesting}, mutation testing has been studied extensively. Numerous mutation tools have also been proposed for different programming languages like  C~\cite{c-mutation},
Java~\cite{java-mutation}, C\#~\cite{csharp-mutation}, Fortran~\cite{fortran-mutation}, SQL~\cite{sql-mutation}. Mutation on intermediate representation (IR) has been studied as well \cite{mull}. In \cite{hariri2019comparing} authors compared source-level and IR-level mutation testing and observed closely correlated mutation scores. 

Mutation testing relies on two fundamental hypotheses: Competent  Programmer  Hypothesis~\cite{acree1979mutation}, and Coupling  Effect  hypothesis~\cite{coupling-fact}. The former hypothesis states that the programmer tends to develop the program with minimal faults and close to the correct. Therefore, a mutation analysis with simple mutation operators still can simulate actual faults that are introduced by programmer.
The latter hypothesis states that complex program faults can be de-coupled into simpler ones. Therefore a test that catches simple faults can catch a high percentage of complex ones, where complex faults are associated to making more than one change into the original program.  
The mutation operators that we used in this paper are based on the ones in \cite{hariri2018srciror} which target four main categories: conditional instructions, logical instruction, arithmetic instructions, and constant values.  
We explain these mutation operators in more details in \autoref{sec:mut-operators}.

Offutt et al~\cite{offutt1997automatically} formulated the conditions that a test suite can kill a mutant. First of all, the mutated instruction should be \textit{reachable} by at least one test input in the test suite. Second, the reached mutated instruction should enter the execution to an incorrect state. Third, the incorrect state should propagate and reach the program output. 

Depending on the mutation operators and level of mutation application, the number of generated mutants varies. Hariri et al~\cite{hariri2019comparing} reported that source-level mutation produced fewer mutants compared to IR-level mutation. Our experience is in alignment with this finding where binary level mutation may create numerous mutants. Not all the generated mutants have same quality in evaluating test adequacy. In this paper we filter out the \textit{trivial} mutants which are the ones that fail to execute on any input data. Such mutants has no contribution in evaluating the effectiveness of the test suite.

\subsection{Binary Rewriting} \label{sec:bin-rewrite}
In this section, we will go through the publicly available binary rewriting solutions and compare their approach with regard to structural recovery, data type extraction, and limitations on supported architectures. Binary rewriting is the process of modifying a compiled program in such a way that it remains executable and functional without accessing  the source-code. Binary Mutation is the process of purposefully introducing faults into a program without having the source-code. While in the definition of that there is not any criteria on soundness of the resulting binary, for the purpose of test suite evaluation we require the injected fault passes some trivial tests.

There are two methods of modifying the binary, one statically and the other over the span of a program execution which is called dynamic. In static approach, we keep a copy of modifications on a separate file on disk. However, dynamic rewriting which is denoted as instrumentation applies the modifications at the runtime. 

Dynamic rewriters like PIN \cite{luk2005pin} and DynamoRio \cite{bruening2012transparent} apply instrumentation at defined locations in memory. Compared to static rewriter, at runtime there is only one code block to translate and deal with. However, depending on the implementation, frequent control flow changes or context switches between rewriter built-in virtual machine (VM) and operating system adds up overhead. Another kind of dynamic rewriting has been introduced by dynamic translating of program to an IR \cite{becker2012binary}. Tools like Valgrind or QEMU can lift the binary to an IR and perform code mutations on top of the generated IR. Then, they can translate it back to apply those modifications during execution. While this approach seems to be effective, it suffers from runtime overhead. 

Our focus is on static rewriting schemes and hence we dedicate the rest of this section to it. There are three known static rewriting schemes. The oldest one is based on \textit{detouring} at assembly level. \textit{Detouring} works by hooking out the underlying instruction. There are two flavors of detouring technique, \textit{patch-based instrumentation} and \textit{replica-based instrumentation} \cite{wenzl2019hack}. \textit{Patch-based instrumentation} replaces the instruction with an unconditional branch into a new section containing instrumentation, replaced instruction, and a control flow transfer back to the patch point. Detouring is a direct rewriting and is ISA dependent which makes the approach inconvenient. This approach introduces a high performance degradation given the two control transfers at patch points. 

\textit{Replica-based instrumentation} method places jump instructions at control flow changing destinations to a replicated code section containing both a copy of the original code and instrumentation. All memory references in this section are modified to maintain less control flow transfers between original and replicated section. While the performance of this approach is better compared to the \textit{patch-based}, the size of the resulting binary is noticeably increased. \cite{ahmadi2021mimosa} incorporated replica-based instrumentation in the malware domain by hooking APIs often used by malware authors to detect analysis environment.

\textit{Reassembleable disassembly} is another static rewriting technique which works by recovering relocatable assembly code. Hence, instrumentation could be inlined and reassembled back to a working binary. This approach first introduced by UROBOROS~\cite{wang2016uroboros} and then expanded and improved by Ramblr~\cite{wang2017ramblr}. This approach enhances the performance since inlined assembly avoids inserting control flow changing instructions at instrumentation points. As a result, performance penalty caused by jump instructions is alleviated. Ddisasm~\cite{flores2020datalog} is the state-of-the-art tool for reassembling disassembly developed in Datalog and combines novel heuristics in function recovery and data access pattern. 

Among the three options, we chose Ddisasm as our candidate for reconstructing disassembly because of the following reasons: 1) Compared to Ramblr, Ddisasm is about five times faster. 2) UROBOROS scans the data section linearly and considers any machine word-sized buffer whose integer representation falling in a memory region as a memory reference. This assumption under the compiler optimization introduces False-positive and False-negatives \cite{wang2017ramblr}. 3) Ramblr improved the \textit{content classification} by applying strong heuristics like \textit{localized value-set analysis} and \textit{Intra-function data dependence analysis}. To achieve a higher accuracy in Control Flow Graph (CFG) recovery Ramblr heavily relies on using symbolic execution which slows down the rewriting process. Apart from Ramblr heuristics, ddisasm incorporated register value analysis (RVA) as an alternative over traditional Value Set Analysis (VSA). In addition, they introduced Data Access Pattern (DAP) analysis which is a def-use analysis combined with the results of register value analysis for a refined register value inference at any given data access point.

\textit{Full-translation} approach works by translating a low-level machine code to a high-level intermediate representation (IR) using a compiler-based front-end for architecture independent binary rewriting. The process of translating the binary to the IR is called \textit{lifting}, while assembling the IR back to a working executable is denoted as \textit{lowering}. Advantages of lifting the binary to a high-level IR are two folds. First, relying on IR makes the rewriting framework ISA-agnostic, as a result leading to support more architectures. Second, providing the ability to apply program analysis techniques like Value Set Analysis (VSA) \cite{balakrishnan2004analyzing} and optimization passes like Simple Expression Tracker (SET) and Offset Shifted Register Analysis (OSRA) conveniently \cite{di2017rev}. On the other hand, complete translation suffers from changing the structural integrity such as cache locality and CFG. 

Rev.ng \cite{di2017rev} relies on full binary translation by lifting the binary to TCG (the IR used in QEMU) and para-lifting TCG to LLVM-IR to benefit from more advanced transformation and analysis passes for CFG and function boundary recovery. While frameworks like angr \cite{wang2017angr} use lifting to apply more advanced binary analysis on top of the intermediate-level representation, they do not have the functionality of lowering the resulting transformations back to the binary. Moreover, Rev.ng heavily relies on code pointers for identifying function entry points and leverages value-set analysis for a more precise value boundary tracking. \cite{kiaei2020rewrite}

\section{Approach}
\label{sec:approach}

Mutation analysis has been widely studied to evaluate the effectiveness and quality of test suites concerning code coverage and semantic integrity of program. Mutation testing works by replacing operators or operands with a list of candidates inheriting similar behavior. These studies apply mutations directly on source-code or by lifting the code to an intermediate representation (IR) \cite{hariri2019comparing}. In the previous research \cite{emamdoost2019binary}, authors applied Uroboros~\cite{wang2016uroboros}, a static binary rewriter for the purpose of mutation generation. Uroboros recovers program structures from stripped binaries and provide an API for recovering the Control Flow Graph (CFG) and call graphs of a binary. Since Uroboros is capable of recovering the lost relocation information, it is possible to inline assembly in the middle of the binary. While Uroboros provides a relocatable disassembly, we discussed in Section \ref{sec:bin-rewrite} that it is not scalable to real-world examples and lacks correctness and soundness in symbolization process. 

In this paper, we introduce \thename{}, a light-weight scalable binary mutation framework backed with a rich set of mutation operators analogous to source-level mutation engines like SRCIROR \cite{hariri2019comparing}. Relying on binary rewriters which does not work on top of an IR or does not provide any API to work with the intermediate representation, decreases the portability of mutation engine over architectures. \thename{} is a conglomerate of two powerful binary rewriting frameworks, Rev.ng \cite{di2017rev} and Ddisasm \cite{flores2020datalog}. Ddisasm works on top of the Grammatech Intermediate Representation for Binaries (GTIRB) \cite{schulte2019gtirb}, which lifts the binary to an IR and maps it to an assembly representation. Hence, this mapping facilitates the re-assembling of binary. To incorporate the power of LLVM-IR passes for applying more powerful mutations, we integrated Rev.ng with SRCIROR. Rev.ng lifts the binary to Tiny Code Generator (TCG), QEMU's IR, and para-lifts the result to LLVM-IR. In Section \ref{sec:eval}, we evaluate the benefit and challenges of applying these two engines in program mutation domain. In the following, we describe the mutation operators used over the course of our experiments. 

\begin{figure}[h]
    \centering
    \includegraphics[width=.9\linewidth]{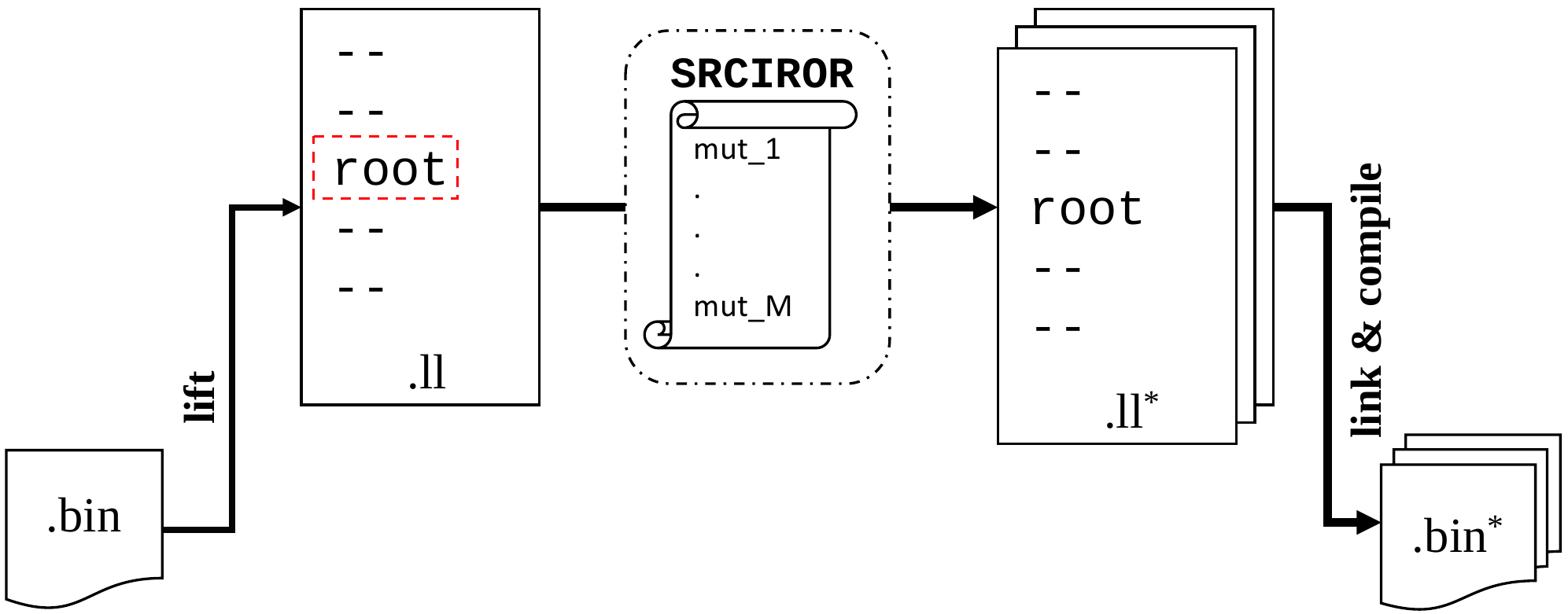}
    \caption{Integration of Rev.ng and SRCIROR to generate mutant binaries}
    \label{fig:mut-revng}
\end{figure}

\subsection{Mutation Operators}
\label{sec:mut-operators}
Numerous mutation generation engines introduced on source-code and to provide the same level of confidence in binary mutation, having an engine close to those is needed. Emamdoost et. al. \cite{emamdoost2019binary} manifested a mutation engine based on Uroboros which only covers conditional branch mutations. To make the assumption close and realistic to errors might occur during development and software patches, they created mutations for either fall-through or taken branch. While control flow modifiers constitute a large amount of effective instructions, we see that there are lots of faults might happen because of Off-by-One (OBO) or integer wrap-around errors which are related to arithmetic operations.

Hence, In our mutant generation, we consider five classes of mutation operators: replacing one arithmetic operator with another different arithmetic operator (AOR), replacing a logic operator with another different logic operator (LCR), replacing the constant operand in an operation with another different constant value (ICR), and replacing the predicate of a comparison with another different predicate (ROR), and skipping instruction by replacing with NOP operator (ISR). We apply only first order mutations, i.e., for each mutated binary only one mutation operator is applied. 

As Table~\ref{tab:mutation-operators} shows, in arithmetic replacements, we substitute every occurrence of an arithmetic operations with an operator from the set of \{+, /, *, \%\}. For logical connectors, we divide them into three sub-categories of logical assignments, logical operators, and bitwise operators of set \{and, or, xor, not\}. For constant replacement we look at the operands passed to an operator and check the immediate values. Then we select a random constant \textit{c} and generate mutants for any combination of \{c+1, c-1, 0, 1, -1, -c\}. For conditional operators, on the assembly level, we either take the next instruction coming after the branch by NOPing the instruction or changing the predicate to take the branch using an unconditional branch. 

To apply the above mentioned mutations, we transfer the target binary to another domain with different abstraction. As discussed in the following sections, we use two different binary rewriters using which we can apply the mutations either at the level of assembly code or at the level of LLVM-IR. Even though our mutation operators are similar in the two tools, the difference in the level of application of mutations leads into different results. More specifically, the mutations applied by Ddisasm have a one-to-one relationship with the mutations in the binary whereas the mutations applied by Rev.ng at the level of LLVM-IR go through the compiling process and therefore may result in modifications to multiple instructions in the binary. Given the optimizations performed on the IR level for Rev.ng size and structure of the resulting binaries are different compared to the original binary. 

\subsection{Rewriting Strategies}

\subsubsection{Ddisasm}
Ddisasm is the main module used for recovering the lost structures and symbolization of unresolved memory addresses. For compatibility with other binary analysis tools, Ddisasm incorporates Google's Protobuf protocol to serialize recovered data and analysis results in GITRB. One distinction between GTIRB and other IRs like VEX, LLVM-IR \cite{lattner2004llvm}, and BAP's BIL \cite{brumley2011bap} is that mainly IRs used to represent the semantics of assembly instruction with more verbosity level to include architecture modifications to registers and CPU flags. While, the main idea behind GTIRB is to represent the binary structure from a high-level while preserving the assembly content. Simply, GTIRB acts like a container of binary analysis in conjunction with assembly content. 

We show our framework pipeline in Figure~\ref{fig:workflow}. In the beginning, framework takes a binary as input and passes it over to Ddisasm to generate a relocatable disassembly in GTIRB representation. GTIRB contains different abstraction for storing information of \textit{Module}, \textit{Symbols}, \textit{SymbolicExpressions}, \textit{Sections}, \textit{ByteIntervals}. As part of our contribution, we modified the default rewriter in Gtirb-capstone to fix the symbolic expressions for the Code and Data blocks following the insertion points and any reference to shifted symbols in the Auxiliary Data. 

Capstone is an ultimate disassembly engine designed for binary analysis and security research. Unfortunately, for the purpose of our research Capstone had a limited set of instruction grouping which was not sufficient for mutation analysis. We added new set of groups to better cover our analysis. At the moment of writing the paper, Capstone only supported JUMP, CALL, RET, and INTERRUPT groups. We added Arithmatic, Logical, and Bitwise grouping. Next, decoded instructions from \textit{ByteInterval} passed to the Mutation Engine to calculate all the possible first-level mutations based on the instruction category. 

For the purpose of our experiments we randomly sampled 1000 mutations and passed the resulting mutated GTIRB file to Ddisasm-compiler. During this process some of the mutations failed the compilation step due to either missing \textit{.eh\_frame} section or conflicting \textit{.ctor} and \textit{.dtor} sections. Finally, to make sure the mutated binary will not crash because of Segmentation Fault or corrupted heap bins, we pass the mutated binaries to a trivial test. This test basically executes the binary with inputs like \textit{-h} or \textit{-v} to assert the healthy state. 

\subsubsection{Rev.ng}
 
To apply our mutations at the level of LLVM-IR, we chose Rev.ng.
Rev.ng's ability to lift the binary to LLVM-IR provides a few advantages. First, manipulating LLVM-IR is simple and comes with a vast set of open-source tools. Second, applying changes to the IR language of a compiler will not allow generating code that is not compilable hence filters out the unsuitable mutations.

In Figure~\ref{fig:mut-revng}, we show our methodology. We first lift the binary using Rev.ng's lifter. We then apply our mutations from the aforementioned set of mutation operators to the lifted binary. This is implemented in the form of LLVM's compiler optimization passes. We used the SRCIROR's IR mutation setup \cite{hariri2019comparing} for this purpose. Once we have the mutated LLVM-IR of the binary, we run LLVM's linker and compiler to generate the mutated binary. 

From the Rev.ng setup, the \translate{} tool is the main rewriter. \translate{} is able to translate a binary from one target architecture to another. It works by lifting the binary to LLVM-IR using its own lifter and compiling for another architecture.
As a result of lifting, Rev.ng adds extra functions to the lifted LLVM-IR while keeping the code of the original binary inside a function called {\tt \smallf{root}}.
To integrate SRCIROR and Rev.ng, we adjusted the SRCIROR setup for LLVM version 10.0.1 to be consistent with Clang's version used in Rev.ng tools. We also modified it to mutate only the parts from the original binary file, i.e. only the \smallf{root} function.
We then separated the lifting from the link and compile processes in \translate{}. After lifting, we use SRCIROR to enumerate all the possible points and types of mutation with our set of mutation operators. Next, we apply each of the possible mutations one at a time, generating mutated IRs. Finally, we run the linker and compiler processes in \translate{} to generate the mutant binaries.





\section{Evaluation} \label{sec:eval}


\begin{figure}[ht]
    \centering
    \includegraphics[width=\columnwidth]{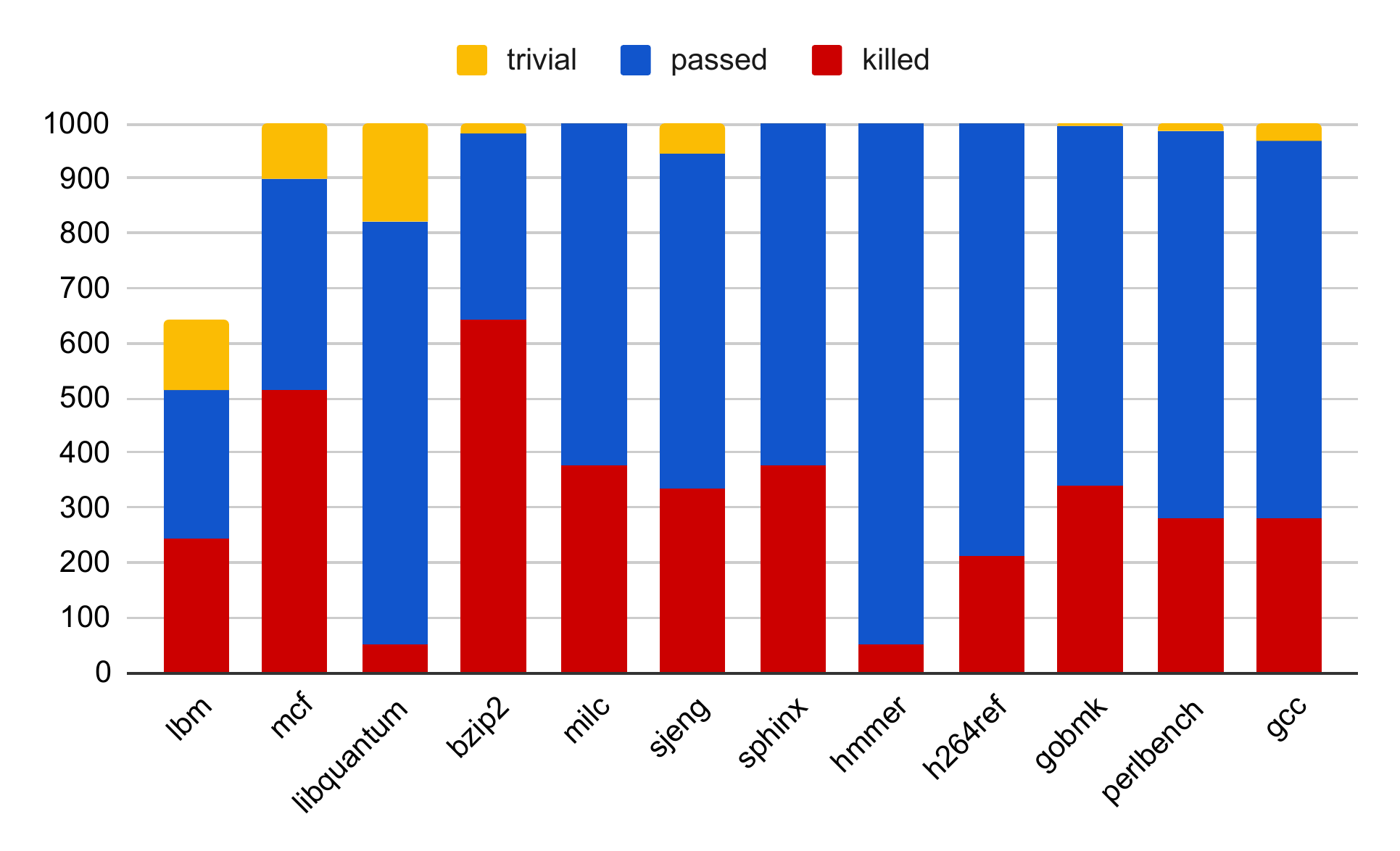}
    \caption{Mutation results for SPEC 2006 benchmark with test input set}
    \label{fig:test-set}
\end{figure}
        
\begin{figure}[ht]
    \centering
    \includegraphics[width=\columnwidth]{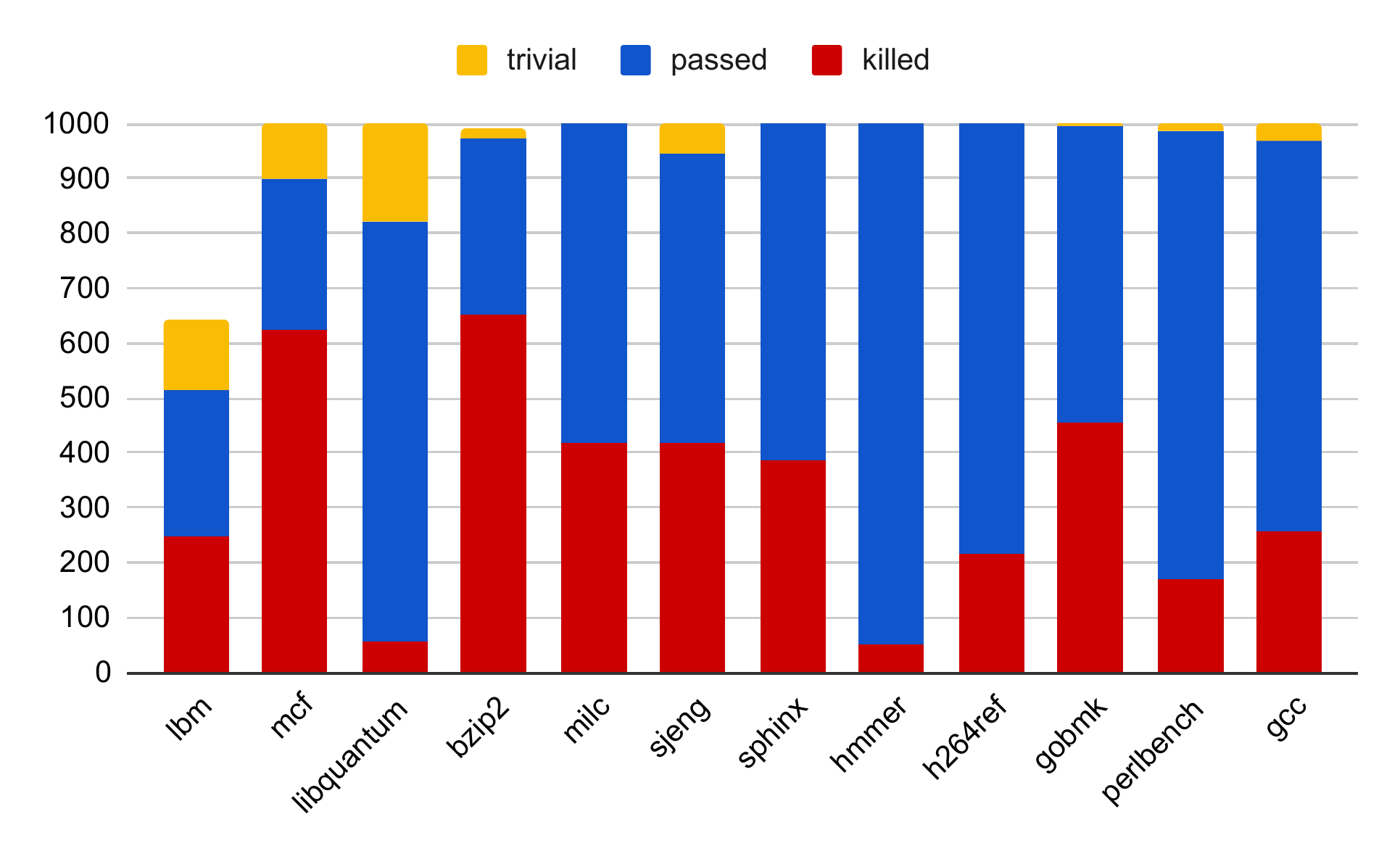}
    \caption{Mutation results for SPEC 2006 benchmark with train input set}
    \label{fig:train-set}
\end{figure}

\begin{figure}[ht]
    \centering
    \includegraphics[width=\columnwidth]{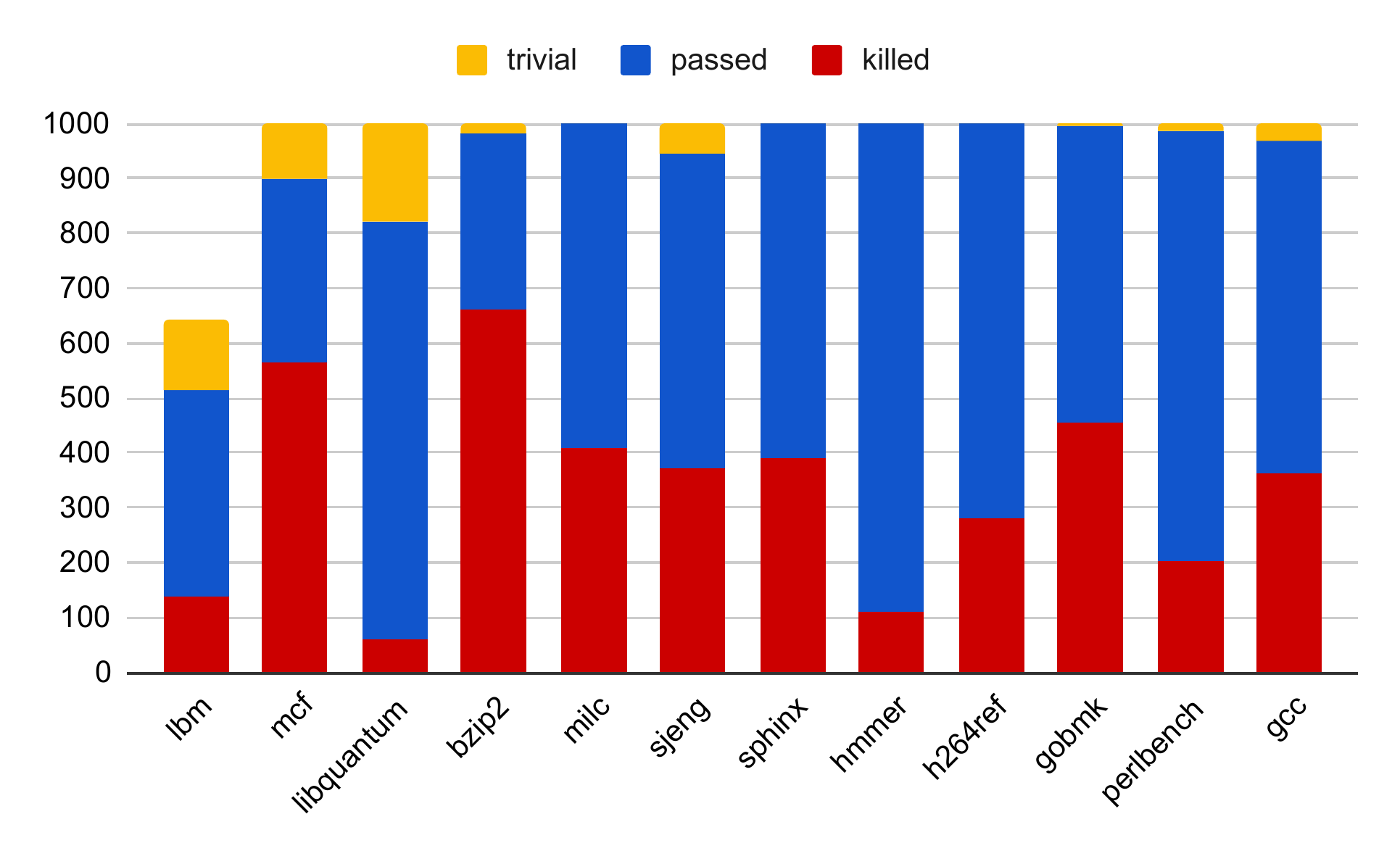}
    \caption{Mutation results for SPEC 2006 benchmark with ref input set}
    \label{fig:ref-set}
\end{figure}

To demonstrate the \thename{}'s scalability and practicality, we used it to generate binary mutants of SPEC CPU 2006 benchmarks compiled for x64 architecture. SPEC CPU is a collection of programs along with three different data sets to evaluate performance of compiler, processor, and or memory. The provided input sets are designed for performance evaluation rather than fault detection or code coverage. That said, the benchmarks are carefully observed to match the expected output. 
Three different input sets mainly differ in size and purposes: the \texttt{test} input is used for checking the benchmark's functionality, \texttt{train} input is used at build time of the benchmark for feedback directed optimizations, and \texttt{ref} input is the largest and actual input data used for performance evaluation.

Based on the approaches described in \autoref{sec:approach} we generate as many as possible mutants, but for the purpose of mutation analysis, we randomly select a subset of 1000 mutants and calculate mutation score per each input data set. \autoref{fig:test-set} shows the ddisasm mutation engine results on the \texttt{test} input set. The experimental results on the \texttt{train} set is presented in \autoref{fig:train-set} and \autoref{fig:ref-set} includes the results of \texttt{ref} set. The proportion of killed mutants for the \texttt{test} input set is relatively lower than for the \texttt{train} and \texttt{ref} input sets. That is mainly because the \texttt{test} input is smaller in size and more simple in a way that does not cover most portions of the code. \texttt{perbench} is an outlier here, as the \texttt{test} input set has kill more mutants compared to other two input sets. That is because based on SPEC CPU documentations, the \texttt{test} input set is derived from the actual test harness shipped with Perl 5.8.7.

\begin{figure*}[h]
\centering
\begin{subfigure}[t]{.5\textwidth}
  \centering
  \includegraphics[width=\linewidth]{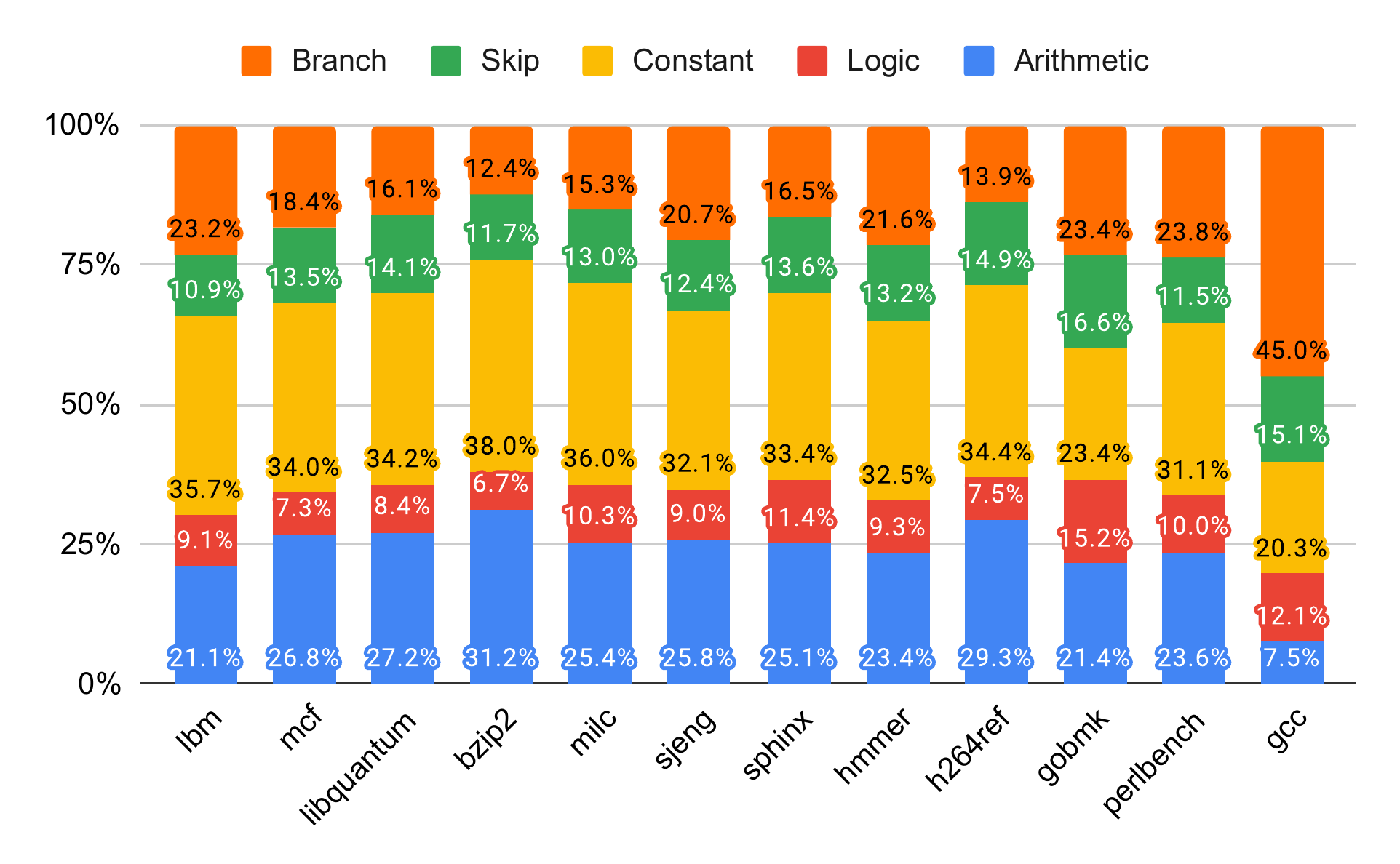}
  \caption{Selected set mutation breakdown.}
  \label{fig:ddisasm-breakdown}
\end{subfigure}%
\begin{subfigure}[t]{.5\textwidth}
  \centering
  \includegraphics[width=\linewidth]{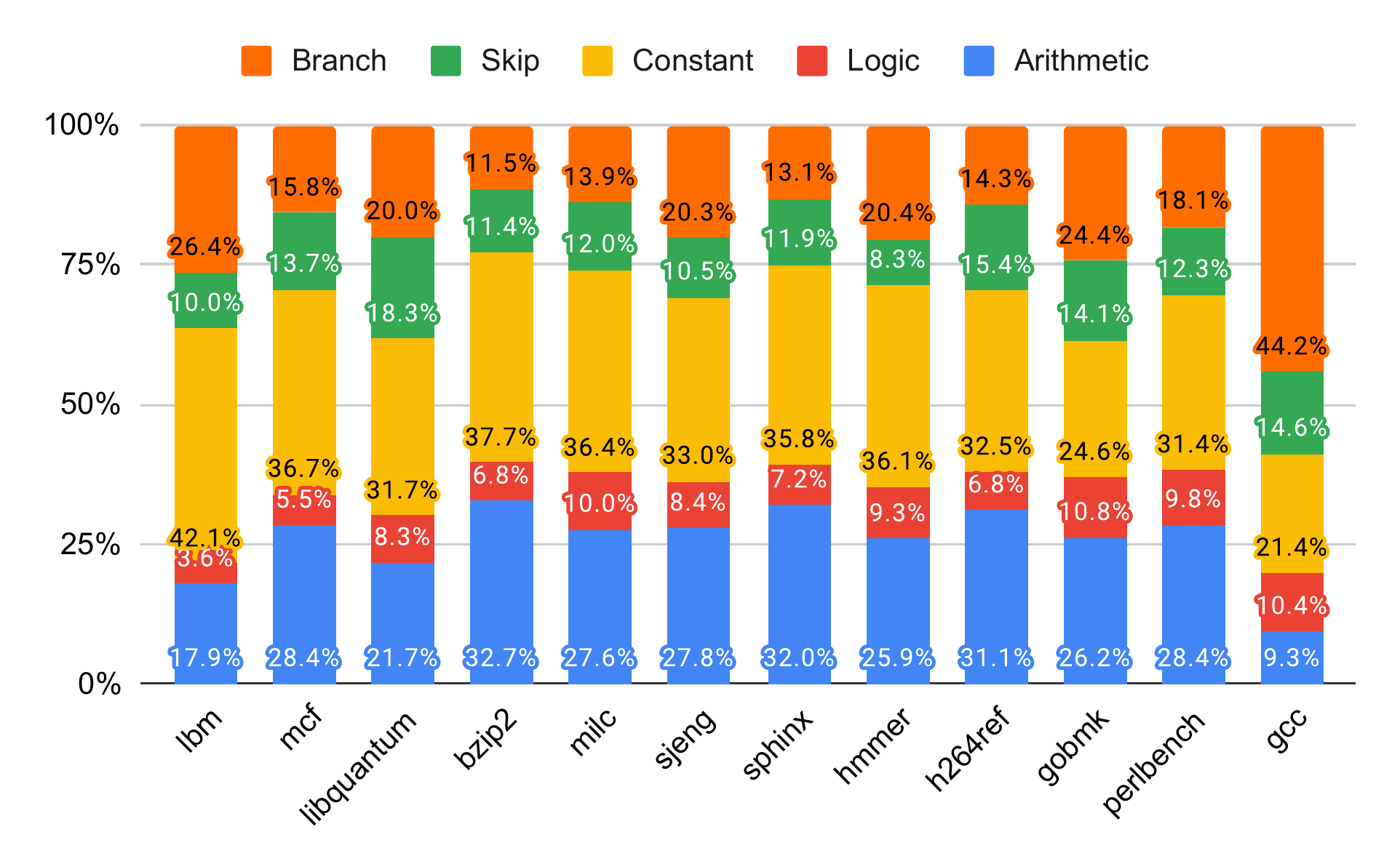}
  \caption{Killed mutants of \texttt{ref} set breakdown.}
  \label{fig:ddisasm-kill-breakdown}
\end{subfigure}
\caption{Mutation Breakdown of 1000 randomly selected mutants over five mutation categories resulting by \texttt{ddisasm} engine. In all of the cases, Arithmetic, NOP (Skip), and Constant operations have greater share compared to others. Greater the size of the binary is the more mutants we are going to expect.} 
\label{fig:test}
\end{figure*}

The number of mutants generated for each benchmark directly correlates with the code size. However, the mutants' diversity in terms of applied mutation operator follows a common pattern among all the benchmarks. Compared to the previous work by Emamdoost et al., we were able to generate more mutants due to our richer mutation operator set. Except for \texttt{lbm}, we could generate more mutants than the sampling rate threshold. While \texttt{lbm} code base only consists of 1500 instructions, we can generate 641 mutants, which means 42\% of instructions fall under at least one mutation category. 

 \begin{table}[h]
 \center \caption{Mutation score of \texttt{test}, \texttt{train}, and \texttt{ref} sets. Mutations score denoted as the number of killed mutants over the total number of mutants. This measure used in assessing the quality of test suite. \label{tab:mut_score}}
\begin{tabular}{llll}
\toprule
\textbf{Binaries}   & \multicolumn{3}{c}{\textbf{Mutation score}}                                                               \\ \midrule
                    & \multicolumn{1}{c}{\textbf{test}} & \multicolumn{1}{c}{\textbf{train}} & \multicolumn{1}{c}{\textbf{ref}} \\ \cmidrule(lr){2-4} 
\textbf{lbm}        & \multicolumn{1}{c}{37.8\%}        & \multicolumn{1}{c}{38.8\%}         & \multicolumn{1}{c}{21.8\%}  \\
\textbf{mcf}        & \multicolumn{1}{c}{51.4\%}        & \multicolumn{1}{c}{60.0\%}         & \multicolumn{1}{c}{56.4\%}       \\
\textbf{libquantum} & 5.2\%                             & 5.7\%                              & 6.0\%                            \\
\textbf{bzip2}      & 64.4\%                            & 65.7\%                             & 66.1\%                           \\
\textbf{milc}       & 37.7\%                            & 41.6\%                             & 40.9\%                           \\
\textbf{sjeng}      & 33.3\%                            & 42.0\%                             & 37.0\%                           \\
\textbf{sphinx}     & 37.7\%                            & 38.4\%                             & 38.8\%                           \\
\textbf{hmmer}      & 5.1\%                             & 5.0\%                              & 10.8\%                           \\
\textbf{h264ref}    & 20.9\%                            & 21.6\%                             & 28.0\%                           \\
\textbf{gobmk}      & 34.0\%                            & 45.5\%                             & 45.5\%                           \\
\textbf{perlbench}  & 27.8\%                            & 16.9\%                             & 20.4\%                           \\
\textbf{gcc}        & 28.2\%                            & 25.5\%                             & 36.4\%      \\
\bottomrule                   
\end{tabular}
\end{table}

We classify mutants in three categories: passed, killed, and trivial based on the initial health-check tests and expected functionality test results. Passed mutants are the ones that generate outputs consistent with original benchmark binaries. Killed mutants are defined as those that fail to generate the same output as the baseline binary because of either calculation errors or falling into an infinite loop. The set of mutants which fail the initial health-check tests or never make the execution due to a segmentation fault or pipe error are classified as trivial. Initial health-check test is a short light-weight input for each binary. For example, in case of \texttt{gcc} is an empty file.  

Infinite loops happen when we apply a control-flow related mutation or changing the logical operator following a comparison. Unbounded loops put the program in a non-returning state. Hence, to cover these edge cases, we applied a timeout chosen twice of the original runtime.

\autoref{fig:ddisasm-breakdown} provides the mutant's breakdown based on the operator used in each mutant generated by \texttt{ddisasm} rewriter. The three most common classes of mutants are arithmetic switching, constant switching, and instruction skip (NOP). This observation constitutes to our initial thought on the diversity of new mutants beyond control flow changing instructions. Dominance of branch mutations in 
 \texttt{gcc} comes from the supremacy of control-flow instructions compared to other binaries which comprises the 22\% of the whole instructions. 
 
 Breakdown of killed mutants on the \texttt{ref} test at \autoref{fig:ddisasm-kill-breakdown} follows a constant pattern complies with \autoref{fig:ddisasm-breakdown}. \autoref{tab:mut_score} manifest the mutation score which is the percentage of mutants killed over the total number of mutants. Compared to \cite{emamdoost2019binary} results, generally we observed higher mutation scores. For example, in \texttt{gcc} they reported a score of less than 15\% for a large code base while ours is 36.4\%. Based on their results, mutation score and binary size has inverse relation. The greater the size of the binary is, the lower mutation score they reported. Since they only relied on conditional branches for mutation, their score for smaller binaries are larger which shows the effectiveness of branch mutation in relatively small binaries. 


Based on \cite{emamdoost2019binary}, they reported problems with compilation flags of \texttt{gcc} and \texttt{gobmk} binaries and their rewriter was limited to a specific version of GCC compiler. In both of our approach, because of powerful binary structure recovery provided by both \texttt{Rev.ng} and \texttt{ddisasm}, we never had such an issue. This shows the portability of our approach compared to their in stable binary mutation of real-world binaries.

\section{Discussion}
\label{sec:discuss}
\begin{table}[t]
\centering
\caption{Size and number of possible mutations on the lifted LLVM-IR by Revng}
\begin{tabular}{lrr}
\hline
\multicolumn{1}{c}{\textbf{benchmark}} & \multicolumn{1}{c}{\textbf{IR size}} & \multicolumn{1}{c}{\textbf{possible mutations}} \\ \hline
lbm                                    & 14MB                                 & 53215                                           \\
mcf                                    & 14MB                                 & 47390                                           \\
libquantum                             & 24MB                                 & 177310                                          \\
bzip2                                  & 30MB                                 & 270217                                          \\
milc                                   & 46MB                                 & 476663                                          \\
sjeng                                  & 53MB                                 & 496435                                          \\
sphinx                                 & 63MB                                 & 699322                                          \\
hmmer                                  & 103MB                                & 1228840                                         \\
h264ref                                & 177MB                                & 2110590                                         \\
perlbench                              & 367MB                                & 4759791                                         \\
gcc                                    & 986MB                                & 13117690                                        \\
gobmk                                  & 283MB                                & 3507198                                         \\ \hline
\end{tabular}
\label{tab:revng-ir-mut}
\end{table}
Mutant generation using the Rev.ng tool had scalability issues that prohibited us from thoroughly analyzing the mutants generated from SPEC CPU 2006. 

Mutant binary size increase becomes a serious issue specifically when the original binary has a moderate code size. Table~\ref{tab:sizeoverhead} shows the size of the rewritten binaries in SPEC 2006 test suite using Rev.ng and Ddisasm.
The rewritten binaries from Rev.ng had a size of from 10$\times$ to 70$\times$ the original binary while Ddisasm rewriter kept the rewritten binaries stable in size. 

\begin{figure}
    \centering
    \includegraphics[width=.9\linewidth]{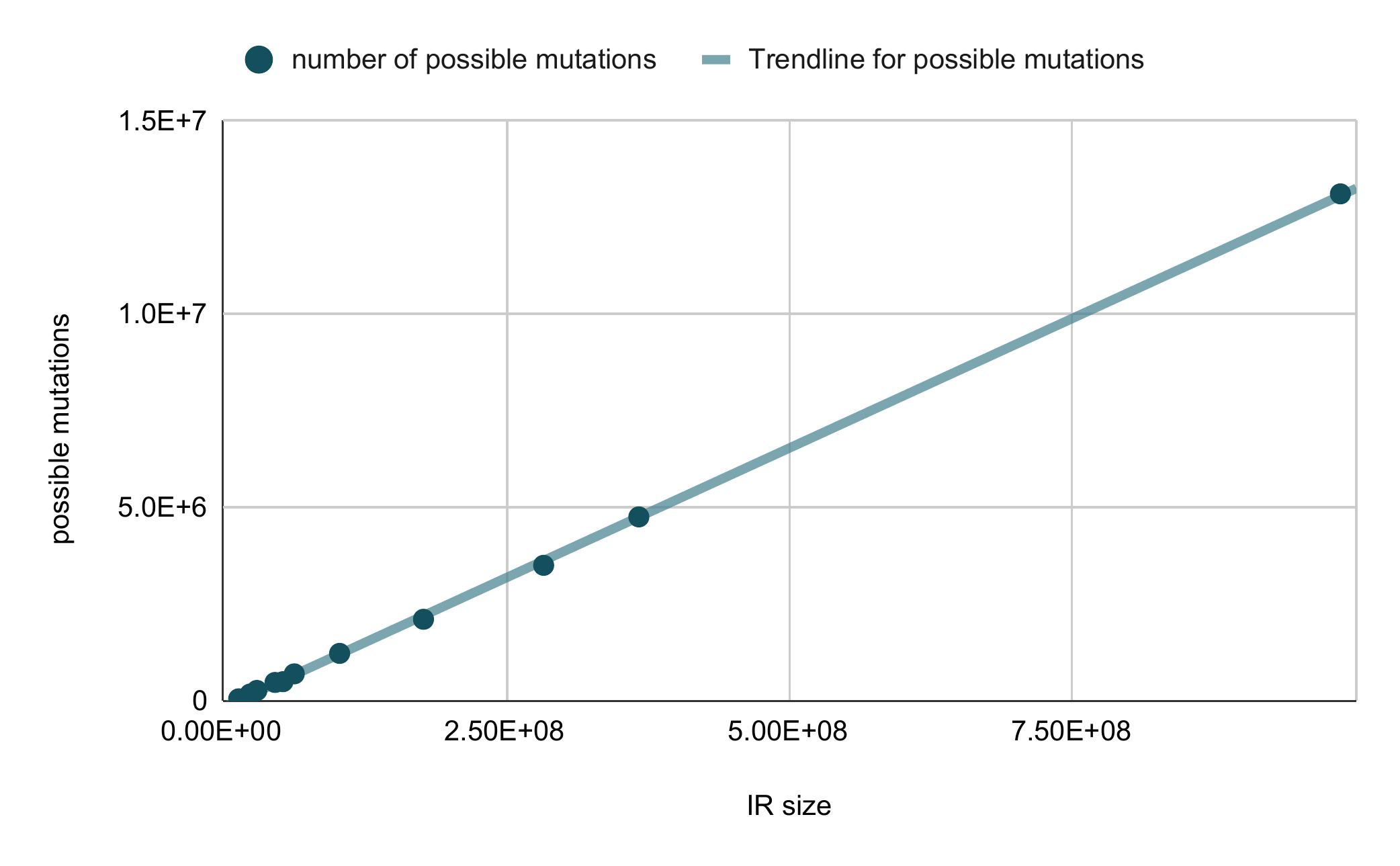}
    \caption{Number of possible mutations by SRCIROR vs. size of the lifted IR by Rev.ng}
    \label{fig:numMutVsIrSize}
\end{figure}

Mutant generation time was another factor that rendered Rev.ng a less appealing rewriting tool for binary mutation. As mentioned before, the number of possible mutants goes beyond millions for binaries with moderate code size like \texttt{gcc}. In such cases the generation time per mutant becomes a decisive factor on practicality and scalability of the binary mutation tool.
Generation of the IR from binaries by Rev.ng took less than a minute for small binaries such as {\tt lbm} and {\tt mcf} but about 20 minutes for {\tt perlbench} and 30 minutes for lifting {\tt gcc}.
Furthermore, the generated IR itself is of significant size as pointed out in Table~\ref{tab:revng-ir-mut}. The bulky size of the lifted IR made the total number of applicable mutations very large. As Figure~\ref{fig:numMutVsIrSize} shows, number of possible mutations by SRCIROR is linearly related to the size of its IR input with the trendline of $0.0134 x - 172510$ and coefficient of determination of $R^2=1$. 

\begin{table*}[t]
\centering
\caption{Overhead in code size resulted from Revng and Ddisasm binary rewriters for SPEC 2006 binaries}
\begin{tabular}{lrrrrr}
\hline
\multicolumn{1}{c}{\multirow{2}{*}{\textbf{benchmark}}} & \multicolumn{1}{c}{\multirow{2}{*}{\textbf{original size}}} & \multicolumn{2}{c}{\textbf{Revng}}                                                          & \multicolumn{2}{c}{\textbf{Ddisasm}}                                                        \\ \cline{3-6} 
\multicolumn{1}{c}{}                                    & \multicolumn{1}{c}{}                                        & \multicolumn{1}{c}{\textbf{rewritten size}} & \multicolumn{1}{c}{\textbf{rewrite overhead}} & \multicolumn{1}{c}{\textbf{rewritten size}} & \multicolumn{1}{c}{\textbf{rewrite overhead}} \\ \hline
lbm                                                     & 22kB                                                        & 1.2MB                                       & 54.5$\times$                                  & 22kB                                        & 1$\times$                                             \\
mcf                                                     & 23kB                                                        & 1.2MB                                       & 52.2$\times$                                          & 22kB                                        & 0.96$\times$                                          \\
libquantum                                              & 51kB                                                        & 3.5MB                                       & 68.6$\times$                                          & 46kB                                        & 0.9$\times$                                           \\
bzip2                                                   & 69kB                                                        & 4,1MB                                       & 59.4$\times$                                          & 68kB                                        & 0.99$\times$                                          \\
milc                                                    & 142kB                                                       & 5.7MB                                       & 40.1$\times$                                          & 134kB                                       & 0.94$\times$                                          \\
sjeng                                                   & 154kB                                                       & 8.1MB                                       & 52.6$\times$                                          & 149kB                                       & 0.97$\times$                                          \\
sphinx                                                  & 198kB                                                       & 7.5MB                                       & 37.9$\times$                                          & 196kB                                       & 0.99$\times$                                          \\
hmmer                                                   & 314kB                                                       & 13MB                                        & 41.4$\times$                                          & 308kB                                       & 0.98$\times$                                          \\
h264ref                                                 & 566kB                                                       & 22MB                                        & 38.9$\times$                                          & 552kB                                       & 0.98$\times$                                          \\
perlbench                                               & 1.2MB                                                       & 47MB                                        & 39.2$\times$                                          & 1.5MB                                       & 1.25$\times$                                          \\
gcc                                                     & 3.6MB                                                       & 127MB                                       & 35.3$\times$                                          & 3.5MB                                       & 0.97$\times$                                          \\
gobmk                                                   & 3.9MB                                                       & 39MB                                        & 10.0$\times$                                          & 4.3MB                                       & 1.1$\times$                                           \\ \hline
\end{tabular}
\label{tab:sizeoverhead}
\end{table*}

\section{Future Work}
In this research, we addressed the limitations of binary mutation by employing more robust binary rewriting approaches and also, adopting a richer set of mutation operators inspired by  source-level mutation. However, we did not explore the possible vulnerabilities that could be introduced by each of those mutations that resemble a real-bug in a program. Such observations can be helpful for example to further investigate which mutation works better for proof-testing patches in a binary. For example to generate mutants resembling double fetch vulnerabilities
by performing a light-weight def-use chain analysis on the operators working on same memory locations. One possible mistake  causing this issue is the incorrect replacement of lines of code during a patch. While a manual test-case might short fall to test the patch comprehensively, a good binary mutation engine can test for these corner cases. 

\section{Conclusion}
\label{sec:conclude}
In this paper, we proposed \thename{}, a new binary mutation tool. \thename{}'s modular design allowed us to adapt and compare the two latest binary rewriting tools: Ddisasm and Rev.ng. Based on our implementation and evaluation, Ddisasm proved to be more practical and scalable in terms of both resulting mutant binary size and mutant generation time. \thename{} is able to mutate any binary independent of the original compilation configuration in use.
Compared to the previous works on binary mutation, \thename{} enjoys a richer set of mutation operators. These operators are better representative of real world faults, and thanks to the scalable underlying binary rewriting technique, are easily applicable to any binary program paving the way to generate many mutants. Applying \thename{} to SPEC CPU 2006 benchmarks, we demonstrated its applicability and effectiveness in terms of generating diverse set of mutants. Such diverse mutants evaluated the adequacy of tests for SPEC CPU binaries. 




\bibliographystyle{IEEEtranS}
\bibliography{IEEEabrv,references}
%

\end{document}